\documentclass[pre,twocolumn,superscriptaddress]{revtex4}
\usepackage[utf8]{inputenc}
\usepackage{amsmath}
\usepackage{amssymb}
\usepackage{braket}
\usepackage{comment}
\usepackage{bbold}
\usepackage{graphicx,color}
\DeclareMathOperator{\Tr}{Tr}

\def\be{\begin{equation}}
\def\ee#1{\label{#1}\end{equation}}
\def\a{\alpha}
\def\b{\beta}

\def\G{\Gamma}
\def\d{\delta}
\def\D{\Delta}
\def\e{\epsilon}

\def\th{\theta}
\def\Th{\Theta}

\def\m{\mu}

\def\p{\pi}
\def\r{\rho}
\def\s{\sigma}

\def\ps{\psi}

\def\i{\int}

\def\bfe{{\mathbf e}}

\def\bth{\boldsymbol{\th}}

\begin{document}
\begin{abstract} 
The statistics of work performed on a system by a sudden random quench is investigated. Considering systems with finite dimensional Hilbert spaces we model a sudden random quench by randomly choosing elements from a Gaussian unitary ensemble (GUE) consisting of hermitean matrices with identically, Gaussian distributed matrix elements. A probability density function (pdf) of work in terms of initial and final energy distributions is derived and evaluated for a two-level system. Explicit results are obtained for quenches with a sharply given initial Hamiltonian, while the work pdfs for quenches between Hamiltonians from two independent GUEs can only be determined in explicit form in the limits of zero and infinite temperature. 

\end{abstract}


\title{Work distributions for random sudden quantum quenches}
\author{Marcin {\L}obejko}
\affiliation{Institute of Physics, University of Silesia, 40-007 Katowice, Poland}
\affiliation{Silesian Center for Education and Interdisciplinary Research, University of Silesia, 41-500 Chorz{\'o}w, Poland}
\author{Jerzy {\L}uczka}
\affiliation{Institute of Physics, University of Silesia, 40-007 Katowice, Poland}
\affiliation{Silesian Center for Education and Interdisciplinary Research, University of Silesia, 41-500 Chorz{\'o}w, Poland}
\author{Peter Talkner}
\affiliation{Institut f\"ur Physik, Universit\"at Augsburg, Universit\"atsstra{\ss}e 1, 86159 Augsburg, Germany}
\affiliation{Institute of Physics, University of Silesia, 40-007 Katowice, Poland}

\maketitle

\section{Introduction}
The discovery of various fluctuation theorems \cite{BK77,ECM93,GC95,J97,C99} made about 20 years ago has led to quite some theoretical and experimental activity with applications and generalizations in sundry directions both within  classical and quantum physics \cite{BK77,J11,EHM09,CHT11,HT15}. In contrast to fluctuation dissipation theorems \cite{CW51,K57} which quantify the response of the average behavior of an arbitrary system variable on a small perturbation leaving the system close to equilibrium, fluctuation theorems are based on the full statistics of work performed by perturbations that may drive the considered system far away from its initial thermal equilibrium state. 

To the best of our knowledge, an important aspect has not yet been considered: So far, the parameters characterizing the initial and the final Hamiltonians, as well as the full sequence of Hamiltonians that connects them and usually is referred to as the ``force protocol'', have always been assumed as being precisely tuned in an experiment and therefore also considered as exactly known in theoretical studies without allowing for any deviations. In the present work we restrict ourselves to sudden quenches of a system described by an instantaneous change of the Hamiltonian. In this case the force protocol shrinks to two Hamiltonians specifying the initial and the final configurations of the system. Within this scenario we consider the limiting case in which the initial and the final Hamiltonians are statistically independent of each other. 

The paper is organized as follows. In Section \ref{wd} the work probability density function (pdf) for a sudden quench between precisely specified Hamiltonians is reviewed. After a short introduction of the Gaussian unitary ensemble (GUE), 
in Section \ref{RH} we discuss the general form of the work pdf for quenches from deterministic to random and from random to random Hamiltonians in the Subsections \ref{krH} and \ref{rr}, respectively. In Section \ref{TL},  we consider as a particular example random sudden quenches of a two-level system. The paper concludes with Section \ref{C}.
\section{Work distribution} \label{wd}
A sudden quench of a thermally isolated system amounts to an instantaneous change of the system's Hamiltonian from $H_i$ to $H_f$ with eigenvalues $e^\a_m$ and corresponding eigenfunctions $\ps^\a_m$, where $\a =i,f$ refers to the initial and the final Hamiltonians, respectively. For the sake of simplicity we do not allow for degeneracy of the eigenvalues. Under the standard assumption that, within each run of the quench protocol, exactly the same  Hamiltonians are realized, the pdf $p(w)$ to find the work $w$ is given by \cite{CHT11}
\be
p(w) = \sum_{m,n} \d (w -e^f_m +e^i_n) p(m|n) p^i(n)\:,
\ee{pw} 
where 
\be
p(m|n) = |\left (\ps^f_m,\ps^i_n \right )|^2
\ee{pmn}
denotes the quench-induced transition probability between the  states $\ps^i_n$ and $\ps^f_n$. Initially the states are populated with weights $p^i(n) = \left (\ps^i_n, \r^i \ps^i_n \right )$, where $\r^i$ is the initial density matrix.  We  assume the weights $p^i(n)$ to follow a Boltzmann distribution at the inverse temperature $\b$, i.e.                 
\be
p^i(n) = e^{-\b e^i_n} \Big / \sum_{n'} e^{-\b e^i_{n'}}\:.
\ee{pin} 
As a consequence, the Jarzynski equality \cite{J97,K00,T00,TLH07} holds
\be
\langle e^{-\b w } \rangle = e^{-\b \D F}
\ee{JE}
relating the average of the exponentiated work to the free energy difference $\D F =F^f-F^i$ between equilibrium states determined by the final and initial Hamiltonians at the initial inverse temperature $\b$. The free energies are defined in the standard way as
\be
F^\a = - \b^{-1} \ln \sum_k e^{-\b e^\a_k} \quad \a = i,f\:.
\ee{F}
\section{Random Hamiltonians} \label{RH}
We restrict ourselves to systems living in a Hilbert space of finite dimension $N$. 
We first shall take the final Hamiltonian and, in the second case, also the initial Hamiltonian 
from a Gaussian ensemble of hermitean matrices invariant under unitary transformations known as the Gaussian unitary ensemble (GUE) \cite{M91}.
Considering the set of hermitean matrices as an $N^2$-dimensional Euclidean space $\mathcal{E}$ one may introduce the infinitesimal volume element
\be
d H = \prod_n^N d H_{nn} \prod^{N(N-1)/2}_{n<m} d \text{Re}(H_{nm})  d \text{Im}(H_{nm}) \:,      
\ee{dH}
where Re$(z)$ and Im$(z)$ denote the real and imaginary part of a complex number $z$, respectively. Here, the diagonal elements $H_{nn}$ together with the real and imaginary parts of the non-diagonal elements $H_{mn}$ with $n<m$ are Cartesian coordinates spanning the space $\mathcal{E}$.
The probability to find a Hamiltonian $H$ within a region $\mathcal{S} \subset \mathcal{E}$ of the space of all Hamiltonians is given by
\be
\text{Prob}( H \in \mathcal{S}) = \i_{\mathcal{S}} d H \r(H)\:,
\ee{ProbH}
where, for the GUE, the pdf $\r(H)$ has the form \cite{M91}
\be
\r(H)=\frac{1}{(2 \p \s^2)^{N^2/2}} e^{-\frac{1}{2\s^2} (\Tr H^2 - 2 \m \Tr H + N \m^2  )}
\ee{rH} 
with parameters $\m$ and $\s^2$ specifying  the mean values of the diagonal elements and the variances of the diagonal as well as of the real and imaginary parts of the non-diagonal elements, respectively, i.e.
\be
\begin{split}
\m &= \frac{1}{N} \langle \Tr H \rangle =\langle H_{nn} \rangle \quad \text{for all}\ n\\
\s^2&=\frac{1}{N^2} \left [ \langle \Tr H^2 \rangle - \frac{1}{N} \langle \Tr H \rangle^2 \right ]\\
&= \langle (\text{Re}(H_{mn})^2 \rangle =  \langle (\text{Im}(H_{mn})^2 \rangle  \\
&= \langle H_{mm}^2 \rangle -\m^2 \quad\text{for all}\; m\; \text{and}\; n \ne m\:,
\end{split}
\ee{ms} 
while the mean values of the non-diagonal elements vanish
\be
\langle H_{mn} \rangle =0 \quad \text{for all}\;n \ne m\:,
\ee{Hnm}
where the averages $\langle \cdot \rangle = \i dH \cdot \r(H)$ are taken with respect to the GUE pdf (\ref{rH}).

Based on the representation of $H = U H^d U^\dagger$ in terms of the diagonal matrix $H^d_{mn} = e_m \d_{mn}$, and a unitary operator $U = (u_{mn})$ made of the $n^{\text{th}}$ components of the $m^{\text{th}}$ eigenvectors one may introduce the set of eigenvalues $\bfe=(e_1,e_2, \ldots e_N)$ and $N(N-1)$ angles $\bth=(\th_1,\th_2, \ldots \th_{N(N-1)})$  with $\th_\a  \in [0,2\p)$ specifying $U$ as alternative coordinates in the space of Hamiltonians. The infinitesimal volume element then becomes
\be
d H = J(\bfe) d \bfe d \bth \:, 
\ee{dHet}   
where $J(\bfe)$ is the Jacobian of the transformation from the Euclidean coordinates used in (\ref{dH}) to $\bfe$ and $\bth$.
It is given by \cite{M91}
\be
J(\bfe) = \frac{\prod_{n<m} (e_n-e_m)^2}{(2 \p)^N \prod_n^N n!}\:.
\ee{J}
Consequently we find from $\r(H) dH = \r(\bfe, \bth) d\bfe d \bth$ the joint probability $\r(\bfe,\bth)$ of eigenvalues and angles determining the eigenvectors the expression
\be
\r(\bfe,\bth) = \r_\bfe(\bfe) \r_{\bth}(\bth)\:,
\ee{reth}
which factorizes in the pdf $\r_\bfe(\bfe)$ of eigenvalues 
\be
\begin{split}
\r_\bfe(\bfe) &= \frac{1}{(2 \p)^{N/2} \s^{N^2} \prod_n^N n!}\\
& \quad \times \prod_{n<m}(e_n-e_m)^2 e^{-\frac{1}{2 \s^2}\sum_n(e_n-\m)^2}
\end{split}
\ee{rfe}
and the uniform pdf $\r_{\bth}(\bth)$ of the $N(N-1)$ angles 
\be
\r_{\bth} (\bth) =\frac{1}{(2 \p)^{N(N-1)}}\:.
\ee{rth}
Note that the energies $\bfe$ and the angles $\bth$ are statistically independent from each other.

We will consider  two cases where either only the final Hamiltonian is drawn  from a GUE or both are independently taken from generally different GUEs.  

\subsection{Sudden quench from fixed to  random Hamiltonians} \label{krH}
For a sudden quench of a system that initially is described by a precisely known Hamiltonian $H_i$ and ends with a Hamiltonian $H_f$ randomly drawn from a GUE,  the work pdf (\ref{pw}) becomes a random object. Averaging with respect to the realizations of the final Hamiltonian $H_f$ yields the pdf in the form 
\be
\begin{split}
\langle p(w) \rangle_f &= \i d H_f  p(w)  \r_f(H_f)\\
&= \sum_{mn} \langle \d(w-e^f_m+e^i_n) \rangle_{\bfe^f} \langle p(m|n) \rangle_{\bth^f}\: p^i(n)\:,
\end{split}
\ee{pwf}
where $\r_f(H_f)$ is given by the pdf (\ref{rH}) parametrized by constants $\sigma_f$ and $\mu_f$. We have used the fact that the pdf (\ref{reth}) of Hamilton operators factorizes into an energy and an angle part.The part resulting from the energy average is independent of the index $m$ because the joint distribution of eigenenergies $\r_\bfe(\bfe)$ is invariant under arbitrary permutations of the index. Accordingly, we find
\be
\langle \d (w-e^f_m +e^i_n)\rangle_{\bfe^f} = D_f(w+e^i_n)\:,
\ee{Df}
where $\langle \cdot \rangle_{\bfe^f}$ denotes the average with respect to $\r_{\bfe}(\bfe)$ and hence 
\be
D_f(E) = \i d\bfe \d(E-e^f_m) \r_{\bfe^f}(\bfe)
\ee{DE}
is the normalized density of states of the GUE  of the final Hamiltonians. 
The transition probability does only depend on the unitary part of the final Hamiltonian but not on its eigenvalues. Its average over the uniform distribution of angles, $\langle \cdot \rangle_{\bth} = \i d^{N(N-1)} \bth \cdot \r_{\bth}(\bth)$ is invariant under arbitrary index permutations. Consequently the transition probabilities are independent of the final index,
hence yielding
\be
\langle p(m|n) \rangle_{\bth} = \frac{1}{N}.
\ee{pN}  
Putting  Eqs. (\ref{Df}) and (\ref{pN}) into Eq. (\ref{pwf}) we obtain the work pdf in the form 
\be
\langle p(w) \rangle_f = \sum_n D_f(w+e^i_n) p^i(n)\:.
\ee{pwD}
The sum over $m$ in Eq. (\ref{pwf}) yields the factor $N$ which combines with the average transition probability (\ref{pN}) to one. Eq. (\ref{pwD}) presents the first  main result of our work. For systems with a large dimensional Hilbert space the density of states approaches a semi-circle law. Hence the normalized density can be approximated by the expression 
\be
D_f(E) = \frac{1}{2 \p \s^2_f N} \sqrt{4 \s^2_f N -E^2} \:\Th (4 \s^2_f N -E^2)\:,
\ee{Dsc} 
where $\Th(x)=1$ for $x\geq 0$ and $\Th(x)=0$ for $x<0$ denotes the Heaviside function. 
At sufficiently low temperatures, mainly the ground state of the initial Hamiltonian contributes and consequently the work pdf assumes the form of 
the density of states shifted by the ground-state energy of the initial Hamiltonian. For sufficiently large Hilbert space dimension $N$ and low temperatures it is therefore determined by the accordingly shifted semicircle law (\ref{Dsc}).
\subsection{Sudden quench from random to  random Hamiltonians} \label{rr}
We now independently draw the initial and final Hamiltonians 
from GUEs characterized by pdfs $\r_i(H)$ and $\r_f(H)$  in the form (\ref{rH})  with variances $\s^2_i$ and $\s^2_f$, respectively and with 
mean-values differing by an amount $\m=\m_f-\m_i$.  
In such a case, the work pdf reads 
\be
\begin{split}
\langle p(w) \rangle_{i,f} &= \i dH_i dH_f p(w) \r_f(H_f) \r_i(H_i)\\
&=\sum_{m,n} \langle \d(w-e^f_m+e^i_n) p^i(n) \rangle_{\bfe^f,\bfe^i}\\
& \quad \times \langle p(m|n) \rangle_{\bth^f,\bth^i}\:,
\end{split}
\ee{pwif}
where we have taken into account that eigenvalues and angles are statistically independent.
Here, $\langle \cdot \rangle_{\bfe^f,\bfe^i}$ and $\langle \cdot \rangle_{\bth^f,\bth^i}$ denote averages with respect to products of initial and final eigenvalue distributions and angle distributions, respectively.
As before, the average of the transition probabilities generates a uniform distribution, i.e.,
\be
\langle p(m|n) \rangle_{\bth^f,\bth^i} = 1/N\:.
\ee{pmntt}
The energy average can be rewritten by introducing delta-functions with respect to the initial and the final energies as follows
\begin{widetext}
\be
\begin{split}
\langle \d(w-e^f_m+e^i_n) p^i(n) \rangle_{\bfe^f,\bfe^i} &= \i d\e^i d\e^f \langle \d(\e^f-e^f_m) \d(\e^i -e^i_n) \d(w-\e^f + \e^i) p^i(n) \rangle_{\bfe^f,\bfe^i} \\
&=  \frac{1}{N} \i d\e^i d\e^f \d(w-\e^i +\e^f) D_f(\e^f) q_i(\e^i)\:,
\end{split}
\ee{ddd1}
\end{widetext}
where the density of states $D_f(\e^f)$ is defined in Eq. (\ref{DE}). The newly introduced function $q_i(\e^i)$ is the energy pdf of the initial state resulting from the average of the Boltzmann distribution $p^i(n) = e^{-\b e^i_n} Z^{-1}_\b(\bfe^i)$ with respect to the GUE distribution of the initial Hamiltonian. It can be understood as the density of states of the initial Hamiltonian weighted by the initial canonical distribution, $p^i(n) = e^{-\b e^i_n} Z^{-1}_\b(\bfe^i)$. The modified density of states is hence given by 
\be
q_i(\e) = e^{-\b \e} \i d \bfe^i \d(\e - e^i_n) \frac{Z_0}{ Z_\b(\bfe^i)} \r_{\bfe^i}(\bfe^i)\:,
\ee{qi}  
where $Z_0=N$ is the partition function in the limit of infinite temperature ($\b =0$). The modified density is independent of the index $n$ because both $Z_\b(\bfe)$ and $\r_{\bfe}(\bfe)$ are invariant under permutations of the components of $\bfe$. In the limit of infinite temperature the modified density of states approaches the proper density of state of the initial GUE: 
\be  
\begin{split}
\lim_{\b \to 0} q_i(\e) &= D_i(\e)\\
&= \i d\bfe \d(\e -\bfe_n) \r_{\bfe}^i(\bfe)\:.
\end{split}
\ee{qD}
In the low temperature limit,  $ q_i(\e)$  approaches the pdf of the ground-state in the initial GUE.  For large $N$ the ground state is distributed according to a Tracy-Widom law \cite{TW94}.

In conclusion, the work pdf of a quench between two GUEs from an initial state with arbitrary temperature can be expressed as
\be
\langle p(w) \rangle_{i,f} = \i d\e D_f(w+\e) q_i(\e) 
\ee{pwDq}
with $D_f(\e)$ and $q_i(\e)$ defined by (\ref{qD}) and (\ref{qi}). This distribution depends in a non-trivial way on the inverse initial temperature $\b$ and the variances $\s^2_i$ and $\s^2_f$ of the initial and the final GUE, respectively. The dependence of the mean-values $\m_i$ and $\m_f$ can simply be generated from the work pdf for $\m_i=\m_f=0$ by a replacement of the argument $w$ by $w-\m_f+\m_i$. 
Eq. (\ref{pwDq}) is the second main result of our work.
Particular examples of work probabilities are studied below. 

\section{Quench of a two-level atom} \label{TL}
As a simple  example we consider a two-dimensional Hilbert space.
We first study the case of a deterministic initial Hamiltonian which is suddenly replaced by a GUE matrix.    
\subsection{Deterministic to random} \label{TLdr}
\begin{figure}
\includegraphics[width=0.3\textwidth,angle=270]{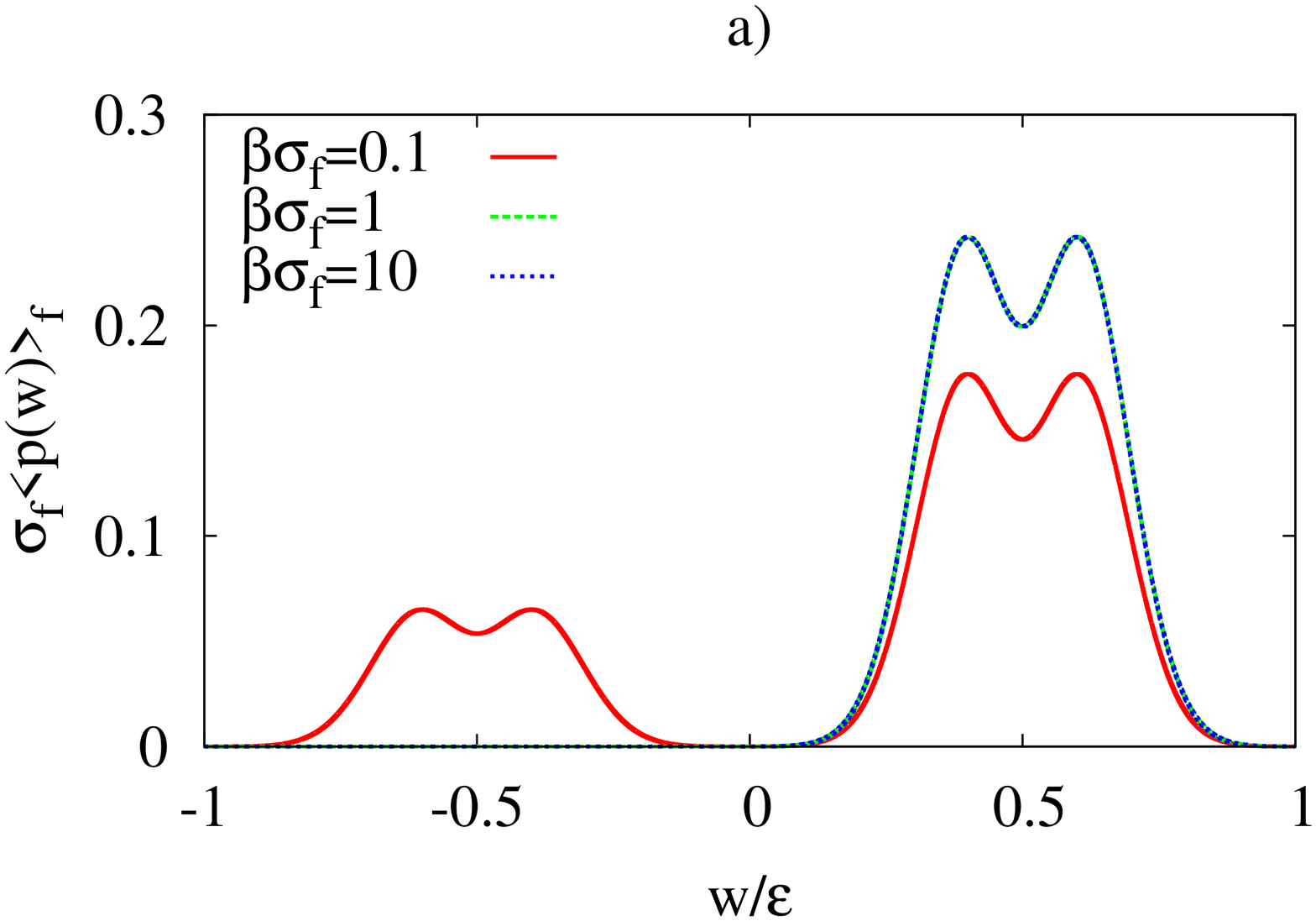}
\includegraphics[width=0.3\textwidth,angle=270]{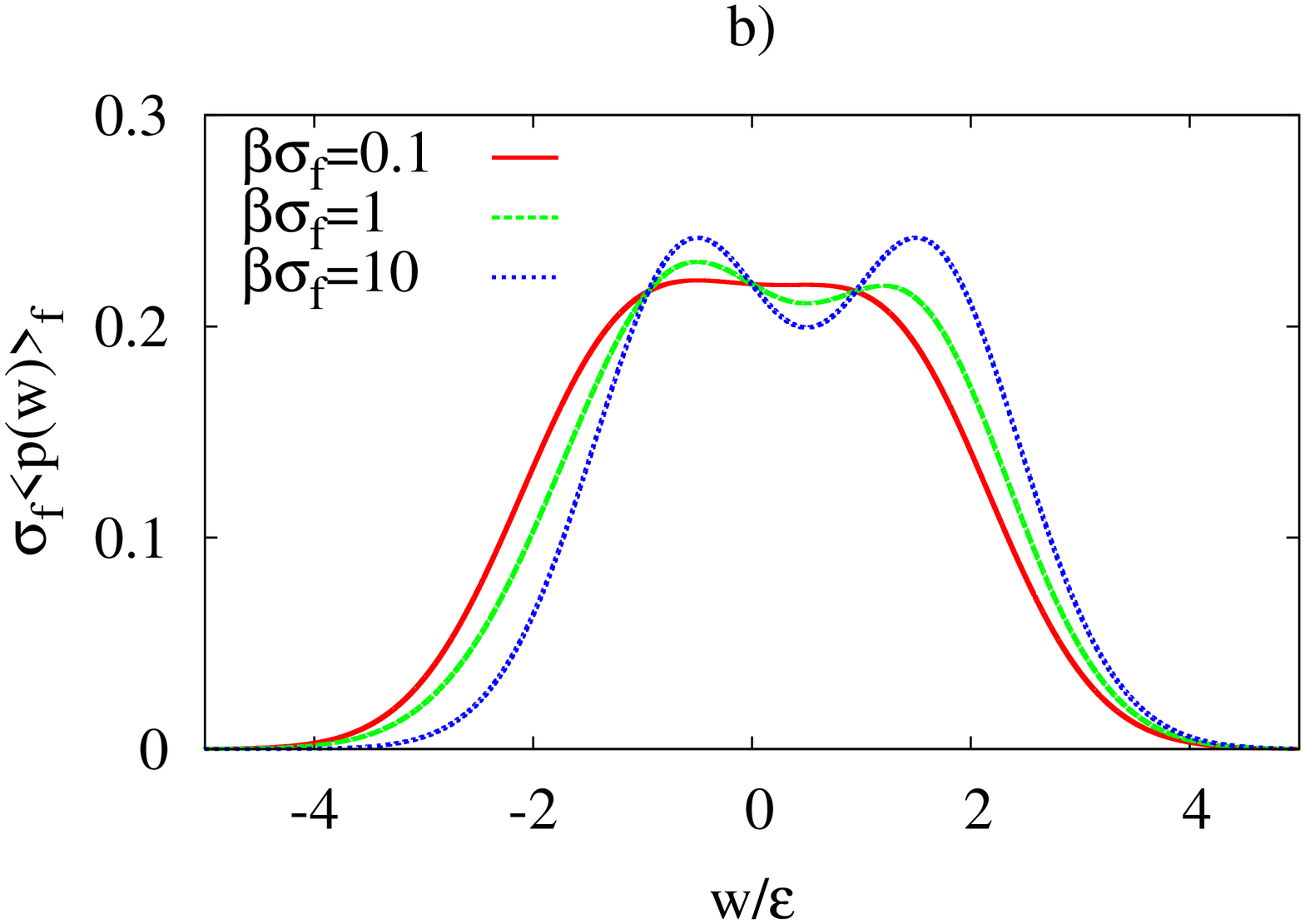}  
\includegraphics[width=0.3\textwidth,angle=270]{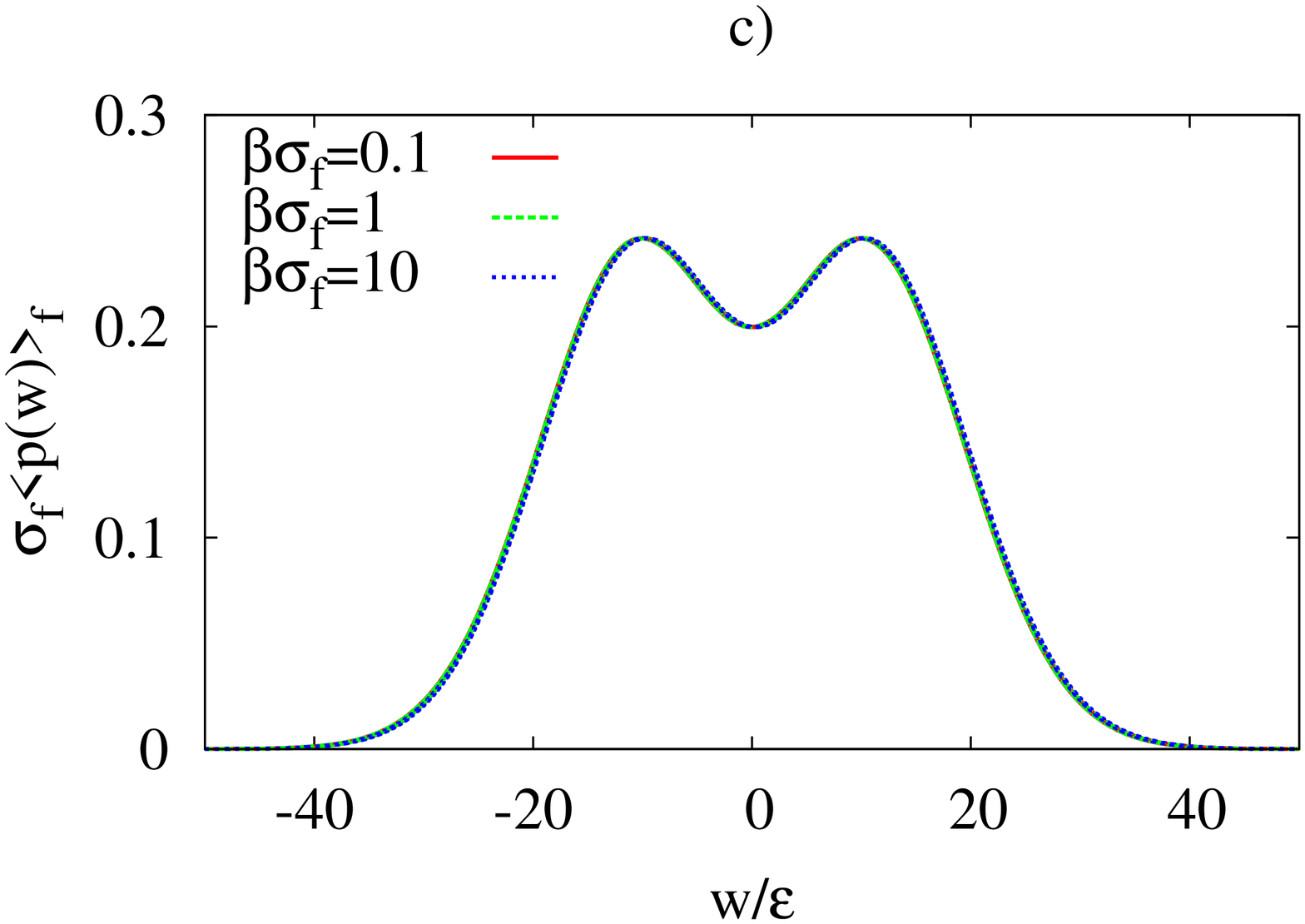}
\caption{The scaled averaged work distribution $\s_f \langle p(w) \rangle^f_{\text{GUE}}$ given by Eq. (\ref{pwf2}) is displayed as a function of $w/\e$ for different inverse temperatures $\b =0.1/\sigma_f$ (red), $\b =1/\sigma_f$ (green) and $\b =10/\sigma_f$ (blue) and different ratios of the initial level spacing and the width of the GUE: $\e/\s_f =10$ in panel (a),  $\e/\s_f =1$ in panel (b), $\e/\s_f =0.1$ in panel (c). With increasing width of the GUE the temperature dependence is less pronounced.}
\label{f1}
\end{figure}

We choose the Hilbert space basis in such a way that the initial Hamiltonian
of the considered two level system takes a diagonal form, i.e.
\be
H_i = \frac{\e}{2} \s_z\:,
\ee{Hi}
where $\e$ is the energy level distance and $\s_z$ is the $z$-component of the Pauli spin matrices $\vec{\s} = (\s_x,\s_y,\s_z)$. The final Hamiltonian which is taken from a GUE can be represented as 
\be
H_f = \vec{\a} \cdot \vec{\s} + h\:,
\ee{Hah}
provided that the components of $\vec{\a} =(\a_x,\a_y,\a_z)$ and the scalar quantity $h$ are independent, Gaussian distributed random variables with $\langle \vec{\a} \rangle_f = 0$, $\langle \a^2_k \rangle_f =  \frac{1}{2} \s^2_f$, $k=(x,y,z)$, $\langle h \rangle_f = \m_f$ and  $\langle (h - \m_f)^2 \rangle_f = \frac{1}{2} \s^2_f$ 
In the sequel we put $\m_f=0$. 
According to Eq. (\ref{rfe}) the joint probability of the eigenvalues $e_1$ and $e_2$ then becomes
\be
\r_f(\bfe) = \frac{1}{4 \p \mathnormal{\s}^4_f } (e_1-e_2)^2 e^{-\frac{1}{2 \s_f^2}(e^2_1 +e^2_2)}.
\ee{rf2}
Using Eq. (\ref{DE}) one obtains for the density of states
\be
D_f(E) = \frac{(E/\s_f)^2 +1}{2 \sqrt{2 \p \s^2_f}} \:e^{- \frac{E^2}{2 \s^2_f} }. 
\ee{Dfe}
The resulting work pdf is
\be
\langle p(w) \rangle_f = D_f(w-\e/2) p + D_f(w+\e/2)(1-p)
\ee{pwf2}
with  $p = 1/(1 + e^{-\b \e})$ 
denoting the ground state population of the initial state. It is displayed for several temperatures and variances $\s^2_f$ in Fig. \ref{f1}. 
For final Hamiltonians from a  GUE with $\m_f \ne 0$ the work $w$ on the right hand side of Eq. (\ref{pwf2}) has to be replaced by $w-\m_f$. 
At low temperatures and for broad distributions, i.e. for large values of $\s_f$ the work distribution approaches the density of states $D_f(E)$ at $E= w +\e/2$
and at $E=w$, respectively, as can be seen from Eq. (\ref{pwf2}).
 
\subsection{Random to random} \label{TLrr}

\begin{figure}[t!]
\includegraphics[width=0.3\textwidth, angle=270]{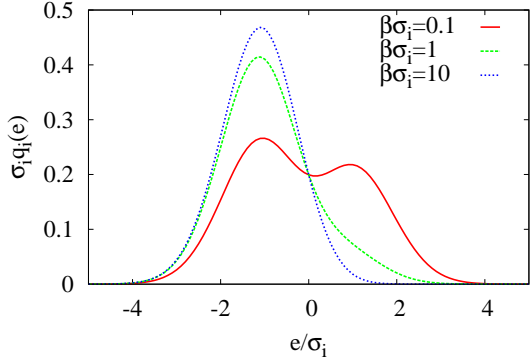}
\caption{The thermally weighted spectral density $q_i(e) \s_i$ is displayed as a function of the scaled energy $e/\s_i$ for different inverse temperatures $\b= 0.1 /\s_i$ (red), $\b= 1/\s_i$ (green) and $\b= 10 /\s_i$ (blue). The distribution for the lowest temperature is already indistinguishable from the ground-state distribution (\ref{rgs}) while at the highest temperature corresponding to $\b \s_i =0.1$ the limiting case of the spectral density $D_i(e)$, which is fully symmetric with respect to $e=0$, is apparently not yet reached.}
\label{qb}
\end{figure}

If the initial Hamiltonian $H_i$ is random from a GUE, from eq. (\ref{qi}) one can get the pdf $q_i(\e)$ in the form 
\be
q_i(\e) = \frac{1}{2 \p \s^4_i} e^{-\frac{1}{2 \s^2_i} \e^2} \i_{-\infty}^\infty d e \frac{(\e -e)^2}{1+e^{\b (\e-e)}} e^{-\frac{1}{2 \s^2_i} e^2}\:.
\ee{qiTL} 
The presence of the partition function in the denominator of the integrand on the right hand side allows one to find analytical expressions only in the limiting cases of high and low temperatures, i.e. for $\b \to 0$ or $\b \to \infty$, respectively.  In Fig. \ref{qb} the scaled and thermally weighted spectral density of states $q(\e) \s_i$ is displayed for different temperatures as a function of the scaled energy $e/\s_i$. It varies from a singly peaked distribution at low temperature to a bimodal distribution at high temperatures.
\begin{figure}[ht!]
\centering
\includegraphics[width=0.3\textwidth, angle=270]{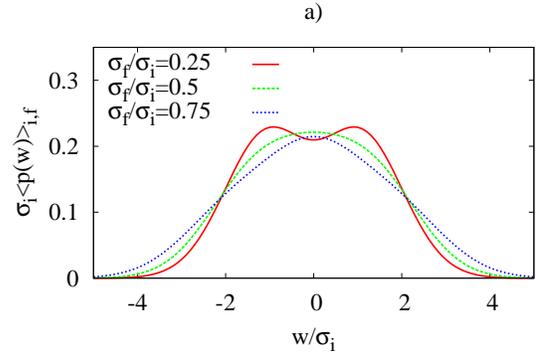}
\includegraphics[width=0.3\textwidth, angle=270]{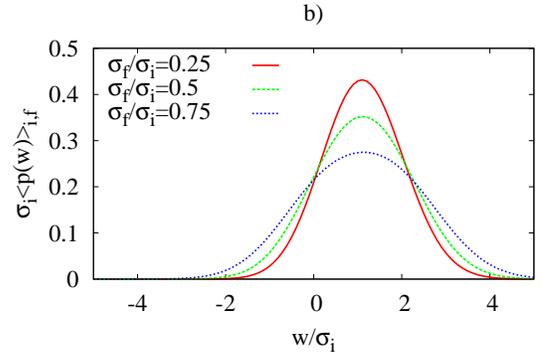}
\caption{The scaled averaged work distribution $\s_i \langle p(w) \rangle_{i,f}$ given by Eq. (\ref{pwDq}) is displayed as a function of $w/\s_i$ for different values of parameters $\s_f/\s_i =0.25$ (red), $\s_f/\s_i =0.5$ (green) and $\s_f/\s_i = 0.75$ (blue). Panel (a) presents the high temperature limit (\ref{pwht}) and panel (b) the zero temperature limit (\ref{pwzt}).}
\label{pdfLim}
\end{figure}
\begin{figure}[ht!]
\centering
\includegraphics[width=0.3\textwidth, angle=270]{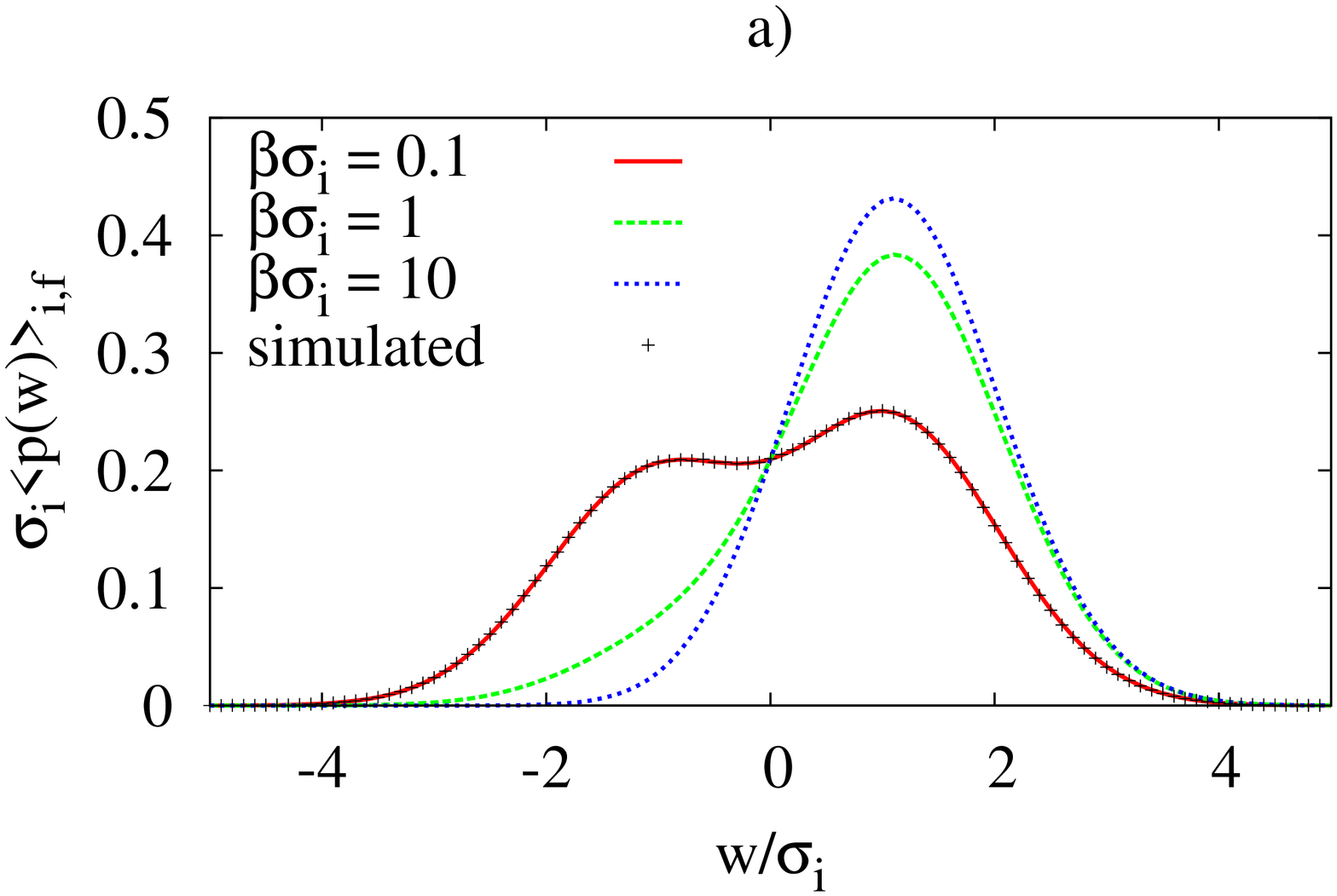}
\includegraphics[width=0.3\textwidth, angle=270]{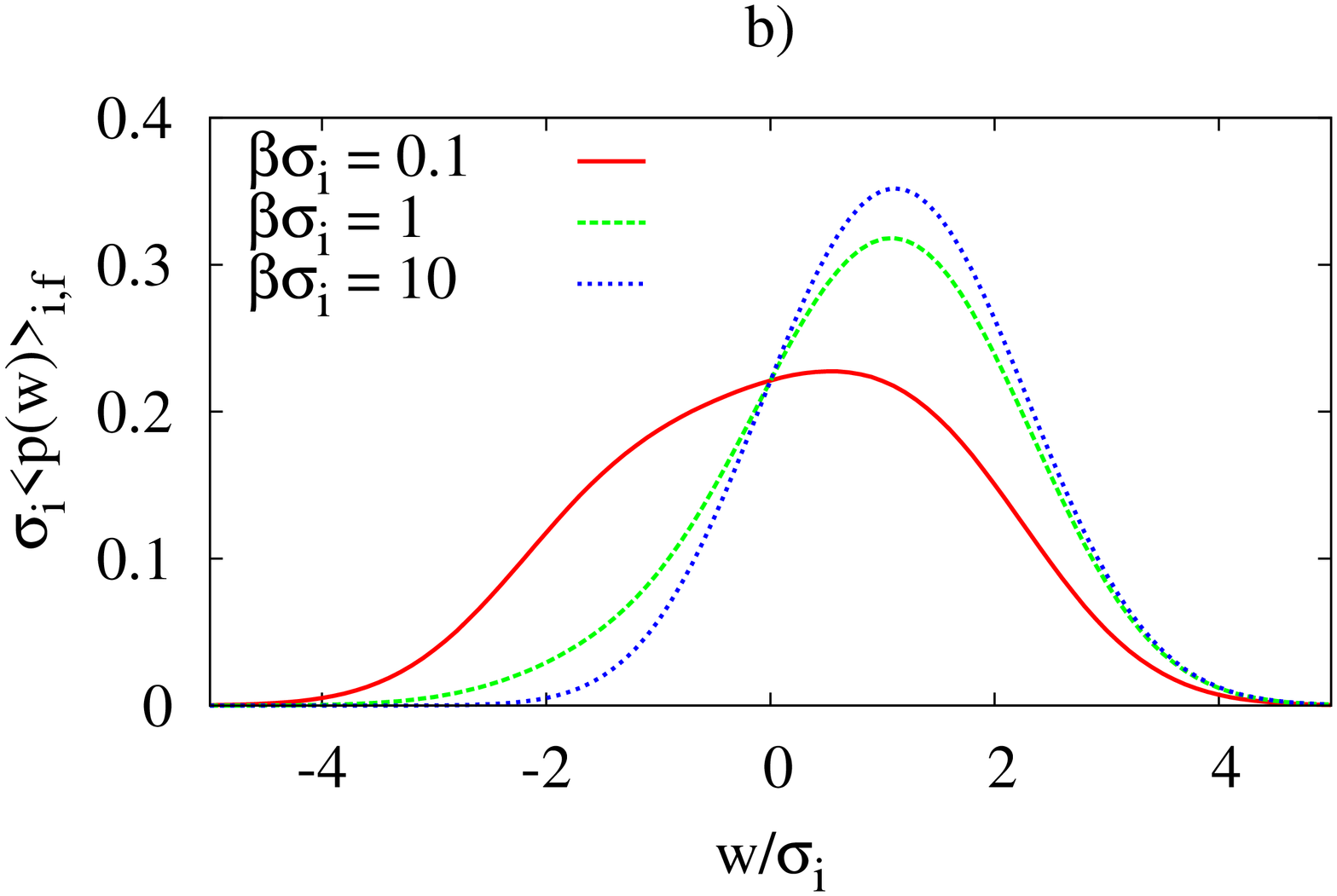}
\includegraphics[width=0.3\textwidth, angle=270]{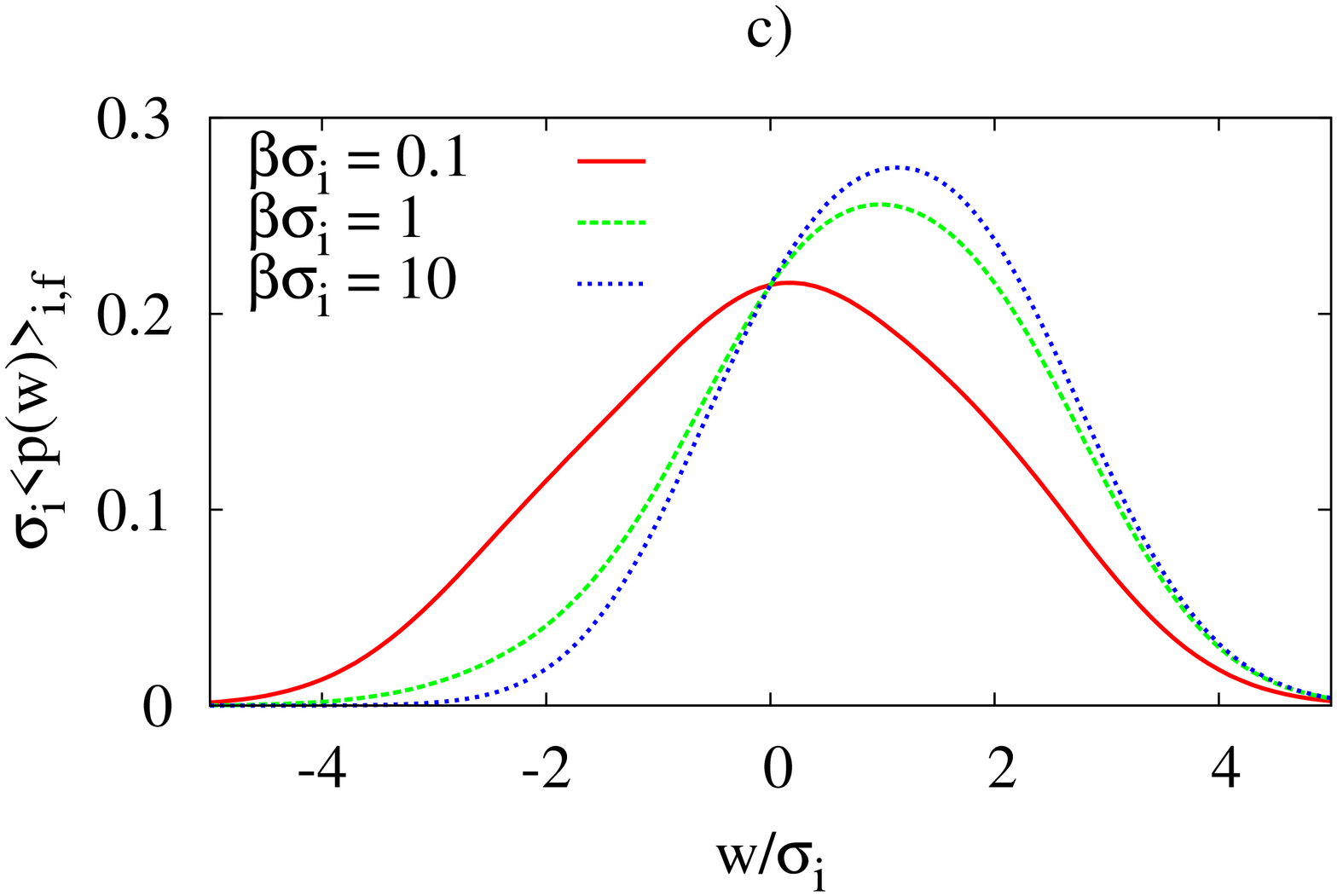}
\caption{The scaled averaged work distribution $\s_i \langle p(w) \rangle_{i,f}$ given by Eq. (\ref{pwDq}) is displayed   as a function of $w/\s_i$ for different inverse temperatures $\b =0.1/\sigma_f$ (red), $\b =1/\sigma_f$ (green) and $\b =10/\sigma_f$ (blue) and different ratios of the GUE widths: $\s_f/\s_i =0.25$ in panel (a),  $\s_f/\s_i =0.5$ in panel (b), $\s_f/\s_i =0.75$ in panel (c). The black crosses in (a) are estimates of the work pdf Eq.(\ref{pw}) at the initial inverse temperature  $\b \s_i =0.1$ from a sample of $10^8$ pairs of energy eigenvalues drawn from two GUEs with $\m_ f=\m_i=0$ and $\s_f/\s_i = 0.25$. The agreement  of the simulation with the numerical integration of Eq. (\ref{pwDq}) with Eqs. (\ref{Dfe}) and (\ref{qiTL}) is excellent.}
\label{pdTemp}
\end{figure}

Using the high temperature limit 
\be
\lim_{\b \to 0} q_i(\e) = D_i(\e)
\ee{leq}
one obtains for  the work pdf (\ref{pwDq}) the expression 
\be
\begin{split}
\lim_{\b \to 0} \langle p(w) \rangle_{i,f} &= \i d\e D_f(\e+w) D_i(\e) \\
&=\frac{P(w)}{4 \sqrt{2 \p \s^2_i}(s^2 +1)^{9/2}} e^{-\frac{w^2}{2 \s^2_i (1 + s^2)}}\:,
\end{split}
\ee{pwht}
where
\be
P(w) =  s^2 \frac{w^4}{\s_i^4}  + 2(1+s^6) \frac{w^2}{\s_i^2} + (1+s^2)^2(2s^4+7s^2+2)\:
\ee{Pw}
with $s = \s_f / \s_i$. Here $D_i(e)$ is the spectral density of the initial GUE. It has the same functional form as $D_f(e)$ given in Eq. (\ref{Dfe}) with $\s_f$ being replaced by $\s_i$.

In the opposite limit of low temperatures the effective pdf of initial energies approaches the ground-state pdf $\r_{gs}(e)$ of the according GUE. In the case of a two-level atom $\r_{gs}(e) = \i d\bfe \d(e- \text{min}(e_1,e_2)) \r_f(\bfe)= 2 \i_0^\infty de_1 \r_f(e_1,e +e_1)$ can be expressed as
\be
\begin{split}
 \lim_{\b \to \infty} q_i(\e) &=\r_{gs}(\e)\\
&= \frac{1}{2\sqrt{2 \p \s^2_i}} e^{-\frac{\e^2}{2 \s^2_i} }\\
&\quad \times \left [ \left (\frac{\e^2}{\s^2_i} +1 \right ) \text{erfc}(\frac{\e}{\sqrt{2} \s_i}) -\sqrt{\frac{2}{\p}} \frac{\e}{\s_i} e^{-\frac{\e^2}{2 \s^2_i}}\right ]
\end{split}
\ee{rgs}
leading to the integral expression for the work pdf:
\be
\begin{split}
\lim_{\b \to \infty} \langle p(w) \rangle_{i,f} = \i d\e D_f(\e+w) \r_{gs}(\e) \:.
\end{split}
\ee{pwzt}
Work pdfs at different parameter values and various ratios of the variances $\s^2_i$ and $\s^2_f$ characterizing the initial and final GUEs are displayed in Fig. \ref{pdfLim}.

\begin{figure}[t!]
\centering
\includegraphics[width=0.3\textwidth, angle=270]{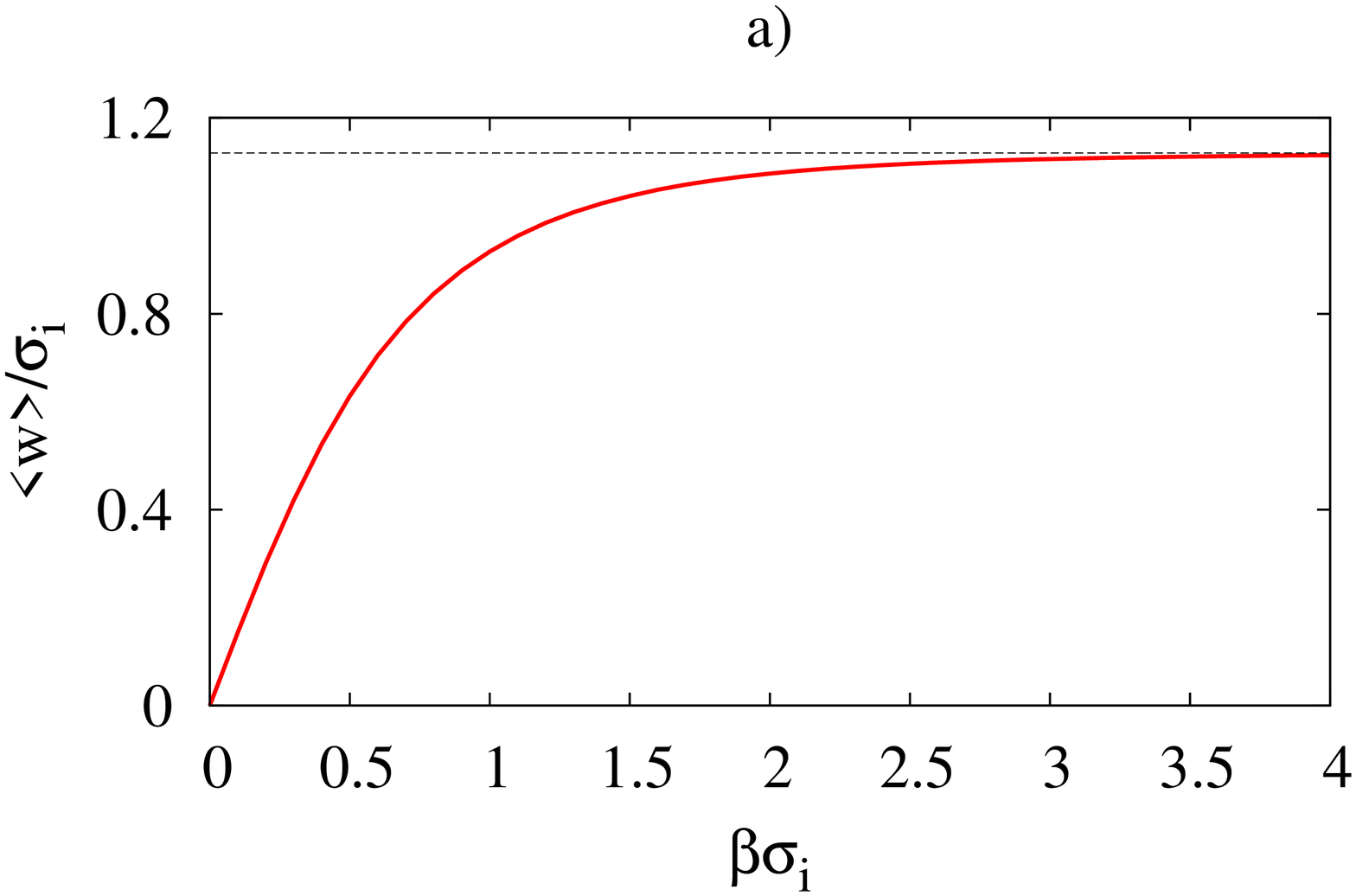}
\includegraphics[width=0.3\textwidth, angle=270]{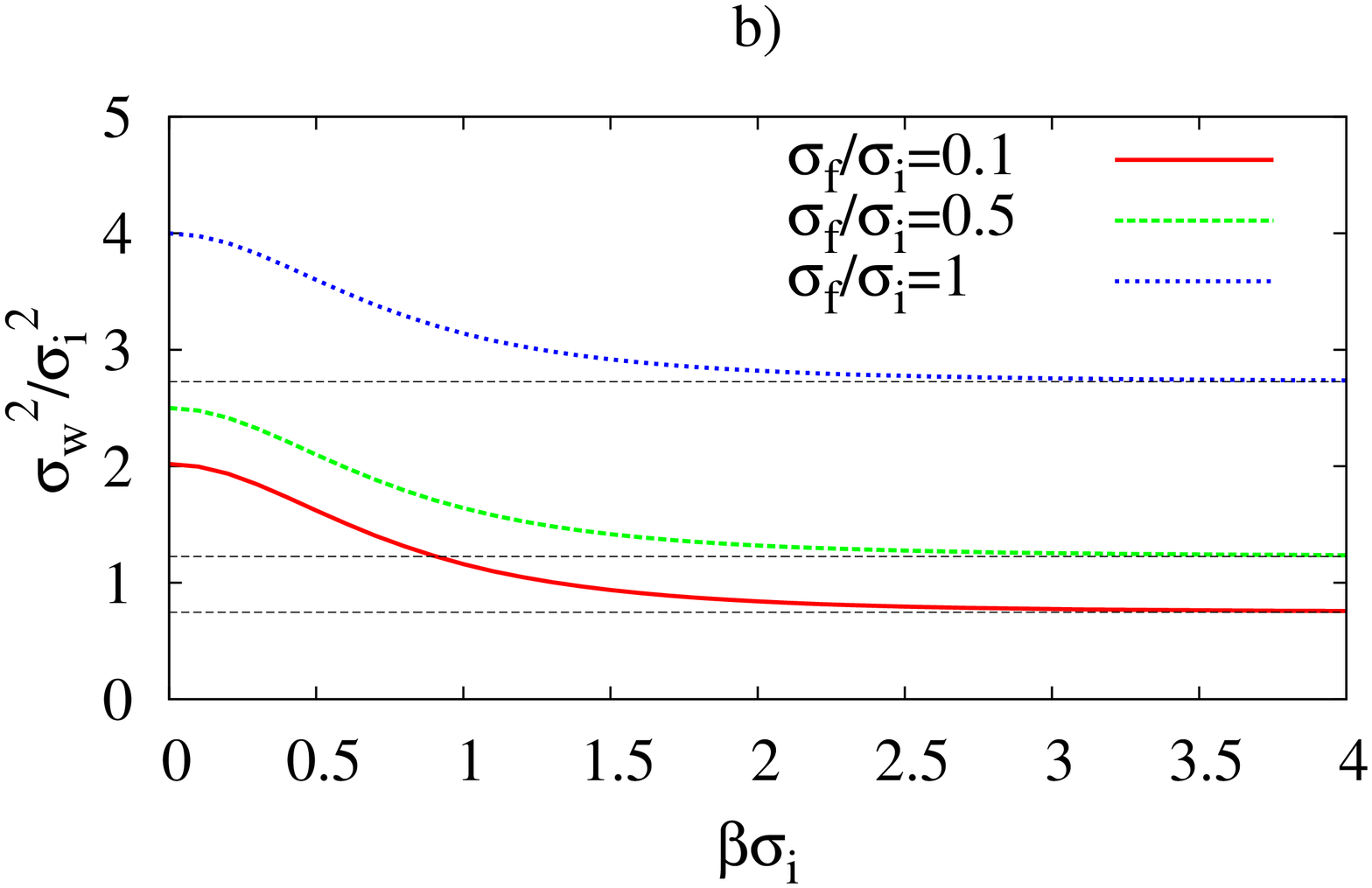}
\caption{Average scaled work $\braket{w} / \s_i$ in panel (a) and variance of the work $\sigma_w^2 /\sigma_i^2$ in panel (b) as  functions of the scaled inverse temperature $\beta \s_i$ for different ratios of the GUE widths: $\s_f/\s_i = 0.25$ (red), $\s_f/\s_i = 0.5$ (green) and $\s_f/\s_i = 0.75$ (blue). Dashed black lines indicate the asymptotic zero temperature limit.}
\label{fig1}
\end{figure}

Finally, we determine the first two moments of work which can be expressed in terms of the moments of the energy density of states $D_f(e)$ and the pdf of initial energies $q_i(\e)$ in the following way
\be
\langle w \rangle = \langle e \rangle^f - \langle e \rangle^i
\ee{avw}
and 
\be
\langle w^2 \rangle =  \langle e^2 \rangle^f -2 \langle e \rangle^f \langle e \rangle^i+ \langle e^2 \rangle^i
\ee{w2}
where
\begin{align}
\label{fav}
\langle e^n \rangle^i &= \i de \, e^n q_i(e), \\
\langle e^n \rangle^f &= \i de \, e^n D_f(e)
\label{iav}
\end{align}
are the $n^{\text{th}}$ moments of initial and final  energies, respectively. 
The corresponding moments of  energies from the thermally weighted initial  GUE can be expressed as
\be
\begin{split}
\langle e \rangle^i &= \frac{\s_i}{4 \sqrt{\p}} \i_{-\infty}^\infty dx \frac{x^3}{1+e^{\b \s_i x}} e^{-x^2/4},\\
\langle e^2 \rangle^i &= 2 \s^2_i.
\end{split}
\ee{im1m2}
For moments of energies drawn from the final GUE one readily finds 
\be
\begin{split}
\langle e \rangle^f &= 0, \\
\langle e^2 \rangle^f&= 2 \s^2_f.
\end{split}
\ee{fm1m2}
While an explicit analytic expression for the first moment is not known, the integration can be performed numerically. The qualitative behavior of the initial energy average is simple: it vanishes for $\b=0$ and decreases with increasing $\b$ towards the zero-temperature asymptote $-2 \s_i/\sqrt{ \p}$. The asymptotic value is well approached for $\b \s_i \gtrapprox 4$.  The second and also all other higher even moments of the initial energy turn out to be  independent of temperature, see the Appendix \ref{APP}.  The resulting first moment and variance $\s^2_w= \langle w^2 \rangle - \langle w \rangle^2$ of the work are displayed in Fig. \ref{fig1} as a function of the scaled inverse temperature $\b \s_i$ for different ratios of the  GUE variances $\s^2_f/\s^2_i$.  

\section{Conclusion} \label{C}
We investigated the statistics of work supplied to a system by a random quench with a final Hamiltonian taken from a GUE. In a single realization the work is determined as the difference between the eigenenergies of the initial and the final Hamiltonians.  
 
Due to fact that the final Hamiltonian is completely uncorrelated from the initial one the transition probabilities between any pair of eigenstates of the initial and final Hamiltonians are identical. Hence, the work distribution is  completely specified by the distributions of eigenenergies corresponding to the initial and final Hamiltonians resulting in a convolution-type expression for the average work pdf. In our setting the distribution of the finally measured energies is determined by the energy density $D_f(E)$ of the respective GUE specified by Eq. (\ref{DE}). For a deterministic initial Hamiltonian the probability with which an energy is detected in the first measurement is determined by the initial state which we assume as being prepared in thermal equilibrium at an inverse temperature $\b$. In those cases in which also the initial Hamiltonian is drawn from a GUE, 
the density of states is weighted by the initial thermal distribution, 
yielding the pdf $q_i(E)$ to find the energy $E$ in the first energy measurement according to Eq. (\ref{qi}). 

As an example we investigated in some detail a two level system suffering a sudden random quench. While the work statistics for a fixed, i.e. non-random, initial Hamiltonian can be completely characterized in terms of an analytic expression for the work pdf, a quench between two random Hamiltonians can be explicitly characterized only in the limiting cases of zero and infinite temperatures. 

Finally we note that much more involved expression must be expected in those physically more realistic cases of quenches that are not completely random but are characterized by Hamiltonians fluctuating about prescribed mean values. In such cases the transition probabilities between the eigenstates of the initial and  final Hamiltonians will become nontrivial such that not only the eigenvalues of the involved matrices contribute to the average work pdf.        

\begin{acknowledgments}
P.T. thanks the Foundation for Polish Science (FNP) for granting him an Alexander von Humboldt Honorary Research Fellowship. The work was supported by the Deutsche Forschungsgemeinschaft via the project DE 1889/1-1 (P.T.) and the NCN Grant No. 2015/19/B/ST2/02856 (J.{\L}.).
\end{acknowledgments}
\appendix
\section{Temperature independence of even work moments} \label{APP}
We consider a random quench of a two level system. Without loss of generality both GUEs out of which the Hamiltonians are chosen are supposed to have vanishing average eigenenergies ($\m_f =\m_i=0$). 
The twice GUE averaged work pdf $\langle p(w) \rangle_{i,f}$ can be formulated in terms of the thermally weighted spectral density of initial states, $q_i(e)$ and the spectral density of final states $D_f(e)$, see Eq. (\ref{pwDq}), reading
\be
 \langle p(w) \rangle_{i,f} = \i de D_f(w+e) q_i(e)\:.
\ee{pw1}
Accordingly, the $2k^{\text{th}}$ moment of the work is given by
\be
\begin{split}
\langle w^{2k} \rangle &= \i dw de\: w^{2k} D_f(w+e) q_i(e)\\ 
&= \i dx de \:(x-e)^{2k} D_f(x)q_i(e)\\
&=\sum_{l=0}^{2k} 
\binom{2k}{l}
\langle e^{2 k-l} \rangle^f \langle e^l \rangle^i\:,
\end{split}
\ee{w2k}
where 
the averages $\langle e^n \rangle^i$ and $\langle e^n \rangle^f$ are defined by  Eqs. (\ref{fav}) and (\ref{iav}), i.e. with respect to the thermally weighted spectral density $q_i(e)$ and the spectral density $D_f(e)$. 
For a two level system the latter are defined by the Eqs. (\ref{qiTL}) and (\ref{Dfe}). 
Note that all odd moments of the final energy vanish,
\be
\langle e^{2 n +1} \rangle^f=0
\ee{e2n1}
and hence only even moments of the initial energy contribute. Those can be evaluated as follows:
\be
\begin{split}
\langle e^{2n} \rangle^i &= \frac{1}{2 \p \s^4_i} \i de de' e^{2n} \frac{(e-e')^2}{1+ e^{\b(e-e')}} e^{-\frac{1}{2 \s^2_i}(e^2 +e'^2)} \\
& =  \frac{1}{4\p \s^4_i} \left ( \frac{1}{2} \right )^{2n} \i du dv (u+v)^{2n} \frac{ u^2}{1+e^{\b u}} e^{-\frac{1}{4 \s^2_i}(u^2+v^2)}\\
& = \frac{1}{4\p \s^4_i} \left ( \frac{1}{2} \right )^{2n} \sum_{k=0}^{2n} 
\binom{2n}{k}
\i_{-\infty}^{\infty} du \frac{u^{2(n+1)-k}}{1+e^{\b u}} e^{-\frac{u^2}{4 \s^2_i}}\\ & \quad \times \i_{-\infty}^{\infty} dv v^k e^{-\frac{v^2}{4 \s^2_i}}\:,
\end{split}
\ee{e2ni}
where we introduced new integration variables $u=e-e'$ and $v=e+e'$ in the second line and expanded the binomial expression under the integrals in order to factorize them. The last integral   
\be
\i_{-\infty}^{\infty} dv v^k e^{-\frac{v^2}{2 \s^2_i}}
\ee{a6}
 vanishes if k is odd. Hence,  only integrals of the form
 \be
I_{2n-k} \equiv \i_{-\infty}^{\infty} du \frac{u^{2n-k}}{1+e^{\b u}} e^{-\frac{u^2}{2 \s^2_i}}
\ee{a7}
  with an even integer exponent $k$ contribute to the sum. The integration range may be split into the negative and the positive $u$-axis to yield:
\be
\begin{split}
I_{2k} &= \overbrace{\i_{-\infty}^0 du \frac{u^{2k}}{1+e^{\b u}} e^{-\frac{u^2}{2 \s^2_i}}}^{u \to -u} +
\i_0^\infty du \frac{u^{2k}}{1+e^{\b u}} e^{-\frac{u^2}{2 \s^2_i}}\\
&= \i_0^\infty du\: u^{2k}\underbrace{\left ( \frac{1}{1+e^{-\b u}} +\frac{1}{1+e^{\b u}} \right )}_{=1} e^{-\frac{u^2}{2 \s^2i}}\\
&= 2^{2n+1} \G(n+\frac{1}{2}) \sigma^{2n+1}_i
\:.
\end{split}
\ee{Ik}
In particular, $I_{2k}$ and therefore all even moments of the initial energy are independent of temperature. Together with the fact that all odd moments of the final energy vanish and the even ones trivially are temperature independent it follows that all even moments of work for a two level system subject to a random  quench are independent of the temperature of the  initial distribution.\\*[10mm]

\end{document}